\documentclass[twocolumn,twoside,slac,floatfix]{revtex4}
\usepackage{graphicx}
\usepackage{epstopdf}
\usepackage{fancyhdr}
\newcommand{\eg}{e.g.}

\newcommand{\etal}{{\it et al. }}

\begin{document}

\title{Terabyte IDE RAID-5 Disk Arrays}

%

\author{D. A. Sanders, L. M. Cremaldi, V. Eschenburg, R. Godang, 
C. N. Lawrence, C. Riley, D. J. Summers}
\affiliation{University of Mississippi, Department of Physics and Astronomy, 
University, MS 38677, USA}
\author{D. L. Petravick}
\affiliation{FNAL, CD-Integrated Systems Development, MS 120, 
P.O. Box 500, Batavia, IL 60510, USA}

\begin{abstract}
High energy physics experiments are currently recording large amounts 
of data and in a few years will be recording prodigious quantities of 
data. New methods must be developed to handle this data and make 
analysis at universities possible.  We examine some techniques that 
exploit recent developments in commodity hardware. We report on tests 
of redundant arrays of integrated drive electronics (IDE) disk drives 
for use in offline high energy physics data analysis. IDE redundant 
array of inexpensive disks (RAID) prices now are less than the cost per 
terabyte of million-dollar tape robots! The arrays can be scaled to 
sizes affordable to institutions without robots and used when fast 
random access at low cost is important.
\end{abstract}

\maketitle

\thispagestyle{fancy}

\section{Introduction}
We report tests, using the Linux operating system, of redundant 
arrays of integrated drive electronics (IDE) disk drives for use in particle 
physics Monte Carlo simulations and data analysis \cite{farm}. Parts 
costs of total systems using commodity IDE disks are now at the \$2000 per 
terabyte level. A revolution is in the making. Disk storage prices have now 
decreased to the point where they are lower than the cost per terabyte of 300 
terabyte Storage Technology tape silos. The disks also offer far better 
granularity; even small institutions can afford to deploy systems. The faster 
random access of disk versus tape is another major advantage. Our tests 
include reports on software redundant arrays of inexpensive disks -- Level 5 
(RAID-5)  systems running under Linux 2.4 using Promise Ultra 133 disk 
controllers that allow disks larger than 137 GB. The 137 GB limit comes from 
28-bit logical block addressing, which allows $2^{28}$ 512 byte blocks on 
IDE disks.   Recently 48-bit logical block addressing has been implemented.
RAID-5 protects data in case of a catastrophic single disk failure by providing 
parity bits. Journaling file systems are used to allow rapid recovery from system 
crashes. 

Our data analysis strategy is to encapsulate data and CPU processing power 
together. Data is stored on many PCs. Analysis of a particular part of a 
data set takes place locally on, or close to, the PC where the data resides. 
The network backbone is only used to put results together. 
If the I/O overhead is moderate and analysis tasks need more than one 
local CPU to plow through data, then each of these disk arrays could be 
used as a local file server to a few computers sharing a local ethernet switch. 
These commodity 8-port gigabit ethernet switches would be combined with a 
single high end, fast backplane switch allowing the connection of a thousand 
PCs. We have also successfully tested using Network File System (NFS) software 
to connect our disk arrays to computers that cannot run Linux 2.4. 

RAID \cite{RAID} stands for Redundant Array of Inexpensive Disks.  Many 
industry offerings meet all of the qualifications except the inexpensive part, 
severely limiting the size of an array for a given budget. This is now changing. 
The different RAID levels can be defined as follow:
\begin{itemize}
\item RAID-0: ``Striped.'' Disks are combined into one physical device where 
reads and writes of data are done in parallel. Access speed is fast but there 
is no redundancy.
\item RAID-1: ``Mirrored.'' Fully redundant, but the size is limited to 
the smallest disk.
\item RAID-4: ``Parity.'' For $N$ disks, 1 disk is used as a parity bit and 
the remaining $N-1$ disks are combined. Protects against a single disk 
failure but access speed is slow since you have to update the parity disk for 
each write. Some, but not all, files may be recoverable if two disks fail.
\item RAID-5: ``Striped-Parity.'' As with RAID-4, the effective size 
is that of $N-1$ disks. However, since the parity information is also 
distributed evenly among the $N$ drives the bottleneck of having to update 
the parity disk for each write is avoided. Protects against a single disk 
failure and the access speed is fast.
\end{itemize}

RAID-5, using enhanced integrated drive electronics (EIDE) disks under Linux 
software, is now available \cite{RAID5}. Redundant disk arrays do provide 
protection in the most likely single disk failure case, that in which a 
single disk simply stops working. This removes a major obstacle to building 
large arrays of EIDE disks. However, RAID-5 does not totally protect against 
other types of disk failures. RAID-5 will offer limited protection in the 
case where a single disk stops working but causes the whole EIDE bus to fail 
(or the whole EIDE controller card to fail), but only temporarily stops them 
from functioning. This would temporarily disable the whole RAID-5 array. 
If replacing the bad disk solves the problem, i.e.~the failure did not 
permanently damage data on other disks, then the RAID-5 array would recover 
normally. 
\begin{table*}[t!]
\begin{center}
\caption{Comparison of Large EIDE Disks for a RAID-5 Array}
\tabcolsep=2.0mm
\begin{tabular}{lcccccr}\hline
Disk Model&Size (GB)&RPM&Cost/GB&GB/platter&Cache Buffer&Warranty\\ 
 \hline
Maxtor D540X \cite{maxtor540}& 160& 5400& \$1.03& 40&2 MB& 3 year\\
&&&&&&\\
Maxtor DiamondMax 16 \cite{maxtor250}& 250& 5400& \$1.09& 83&2 MB& 1 year\\
Maxtor MaXLine\,Plus\,II \cite{maxline}& 250& 7200& \$1.52& 83&8 MB& 3 year\\
Western Digital WD2500JB \cite{WDC250}& 250& 7200& \$1.31& 83&8 MB& 3 year\\
IBM-Hitachi 180GXP \cite{IBM180} & 180 & 7200& \$1.00& 60&8 MB& 3 year\\
\hline
\end{tabular}
\label{Disks}
\end{center}
\end{table*}
Similarly if only the controller card was damaged then replacing 
it would allow the RAID-5 array to recover normally. However, if more 
than one disk was damaged, especially if the file or directory 
structure information was damaged, the entire RAID-5 array would be 
damaged. The remaining failure mode would be for a disk to be delivering 
corrupted data. There is no protection for this inherent to RAID-5; however, 
a longitudinal parity check on the data, such as a checksum record count 
(CRC), could be built into event headers to flag the problem. 
Redundant copies of data that are very hard to recreate are still needed. 
RAID-5 does allow one to ignore backing up data that is only moderately 
hard to recreate.

\section{Large Disks}
In today's marketplace, the cost per terabyte of disks with EIDE interfaces 
is about half that of disks with SCSI (Small Computer System Interface).
The EIDE interface is limited to 2 drives on each bus and SCSI is 
limited to 7 (14 with wide SCSI). The only major drawback of EIDE disks is 
the limit in the length of cable connecting the drives to the drive 
controller. This limit is nominally 18 inches; however, we have successfully 
used 24 inch long cables \cite{star}. 
Therefore, one is limited to about 10 disks per box for an array (or perhaps 20 
with a ``double tower''). To get a large RAID array one needs to use large 
capacity disk drives. There have been some problems with using large disks, 
primarily the maximum addressable size. We have addressed these problems in 
an earlier papers \cite{CHEP98, IEEE}. Because of these concerns and because we 
wanted to put more drives into an array than could be supported by the 
motherboard we opted to use PCI disk controller cards. 
In the past we have tested both Promise Technologies ULTRA 66 and ULTRA 100 
disk controller cards in RAID-5 disk arrays consisting of either 80 or 
100 GB disks\cite{IEEE}. Each of the PCI disk controller cards support four drives. 
We now report on our tests of the Promise Technologies ULTRA 133 
TX2 \cite{Promise133} that supports disk drives with capacity greater than 137 GB.

Using arrays of disk drives, as shown in Table \ref{Disks}, the cost per 
terabyte is similar to that of cost of Storage Technology tape silos. 
However, RAID-5 arrays offer a lot better granularity since they are scalable 
down to a terabyte. For example, if you wanted to store 10 TB of data you 
would still have to pay about \$1,000,000 for the tape silo but only 
\$20,000 for a RAID-5 array. Thus, even small institutions can afford to 
deploy systems. And the Terabyte disk arrays can be used as caches to take 
full advantage of Grid Computing \cite{grid}. 

\section{RAID Arrays}
There exist disk controllers that implement RAID-5 protocols right in the 
controller, for example 3ware's Escalade 7500 series  \cite{3ware}, which will 
handle up to 12 EIDE drives. These controllers cost \$600 and, at the time that 
we built the system shown in Table \ref{Configuration}, 
did not support disk drives larger than 137 Gigabytes \cite{Kent}. Therefore, we focused 
our attention on software RAID-5 implementations \cite{RAID5,HOWTO}, which 
we tested extensively.

There are also various commercial RAID systems that rely on a hardware RAID controller. 
Examples of these are shown in Table \ref{Commercial}. They are typically 3U or larger rack mounted systems.  However, commercial systems have not been off-the-shelf commodity items.  This is changing and the only drawback is that, even allowing for cost of assembly, they are anywhere from twice to over twenty-five times as expensive.
\begin{table}[ht]
\begin{center}
\tabcolsep=1.5mm
\caption[]{Some Commodity Hardware RAID Arrays.}
\label{Commercial}
\begin{tabular}[t]{lrrr}\hline
System &Capacity&Size&Price/GB\,\footnote{Based on suggested 
retail Prices on February 7, 2003\cite{Apple}}  \\
\hline
Apple Xserve RAID & 2.52 TB &3U& \$4.36 \\
Dell EMC CX200 & 2.2 TB &3U& \$13.63 \\
HP 7100 & 2.2 TB &2$\times $3U& \$50.21 \\
IBM DF4000R & 2.2 TB&2$\times $3U& \$20.08 \\
Sun StorEdge T3 & 2.64 TB&3$\times $3.5U& \$54.66 \\
\hline
\end{tabular}
\end{center}
\end{table}
\subsection{Hardware\label{hardware}}
We now report on the use of disks  with capacity greater than 137 GB. The drives we 
consider for use with a RAID-5 array are compared in Table \ref{Disks}. The disk we 
tested was the Maxtor D540X 160 GB disk \cite{maxtor540}. In general, the internal 
I/O speed of a disk is proportional to its rotational speed and increases as a function 
of platter capacity. One should note that the ``spin--up" of these drives takes 1.8-2.5 
Amps at 12 Volts (typically 22 W total for both 12 V and 5V).

When assembling an array we had 
to worry about the ``spin-up'' current draw on the 12V 
part of the power supply. With 8 disks in the array (plus the system 
disk) we would have exceeded the capacity of the power supply that came 
with our tower case, so we decided to add a second off-the-shelf power 
supply rather than buying a more expensive single supply. 
By using 2 power supplies we benefit from 
under loading the supplies. The benefits include both a longer 
lifetime and better cooling since the heat generated is distributed 
over 2 supplies, each with their own cooling fans. 
We used the hardware shown in Table \ref{Configuration} for our array test. 
Many of the components we chose are generic; thus, components from 
other manufacturers also work. 
We have measured the wall power consumption for the whole disk array box in 
Table \ref{Configuration}. It uses 276 watts at startup and 156 watts during 
normal sustained running.

\begin{table}[hbt]
\begin{center}
\caption[]{Components used in our 1\,Terabyte RAID-5 disk array}
\label{Configuration}
\tabcolsep=1.5mm
\renewcommand{\arraystretch}{1.10}
\vspace*{3pt}
\begin{tabular}{lr} \hline
System      & Unit \\
 Component  & Price \\ \hline
40 GB IBM system disk \cite{IBM40}            & \$65 \\
8 -- 160 GB Maxtor RAID-5 disks \cite{maxtor540}      & \$170 \\
2 -- Promise ATA/133 PCI cards \cite{Promise133}     &  \$32 \\
4 -- StarTech 24" ATA/100 cables \cite{star}         &   \$3 \\
AMD Athlon 1.9 GHz/266 CPU \cite{AMD}                & \$77  \\
Asus A7M266 motherboard, audio \cite{asus}           & \$67  \\
2 -- 256 MB DDR PC2100 DIMMs                          & \$33   \\
In-Win Q500P Full Tower Case \cite{inwin}            & \$77   \\
Sparkle 15A @ 12V power supply  \cite{sparkle}       & \$34 \\
2 -- Antec 80mm ball bearing case fans               &  \$8  \\
110 Alert temperature alarm \cite{cool}              & \$15 \\
Pine 8 MB AGP video card \cite{pine}           & \$15 \\
SMC EZ card 10/100 ethernet \cite{smc}       & \$12 \\
Toshiba 16x DVD, 48x CDROM                           & \$36 \\
Sony 1.44 MB floppy drive                            & \$12 \\
KeyTronic 104 key PS/2 keyboard                      & \$7  \\
DEXXA 3 button PS/2 mouse                            & \$4   \\
\cline{2-2}
\hfill Total                                         & \$1922  \\
\hline
\end{tabular}
\end{center}
\end{table}

To install the second power supply we had to modify our tower 
case with a jigsaw and a hand drill. We also had to use a jumper to 
ground the green wire in the 20-pin block ATXPWR connector to fake 
the power-on switch.

When installing the two disk controller cards care 
had to be taken that they did not share interrupts with other highly 
utilized hardware such as the video card and the ethernet card. We 
also tried to make sure that they did not share interrupts with each 
other. There are 16 possible interrupt requests (IRQs) that allow the 
various devices, such as EIDE controllers, video cards, mice, serial, 
and parallel ports, to communicate with the CPU. Most PC operating 
systems allow sharing of IRQs but one would naturally want to avoid 
overburdening any one IRQ. There are also a special class of IRQs 
used by the PCI bus, they are called PCI IRQs (PIRQ). Each PCI card 
slot has 4 interrupt numbers. This means that they share some IRQs 
with the other slots; therefore, we had to juggle the cards we used 
(video, 2 EIDE controllers, and an ethernet). 

When we tried to use a disk as a ``Slave'' on a motherboard 
EIDE bus, we found that it would not run at the full speed of the bus 
and slowed down the access speed of the entire RAID-5 array. This was 
a problem of either the motherboard's basic input/output system (BIOS) 
or EIDE controller. This problem was not in evidence when using the disk 
controller cards. Therefore, we decided that rather than take a factor 
of 10 hit in the access speed we would rather use 8 instead of 9 hard disks. 

\subsection{Software}
For the actual tests we used Linux kernel 2.4.17 with the RedHat 7.2 
(see http://www.redhat.com/) distribution (we had to upgrade the kernel to 
this level) and applied a patch to allow support for greater than 137 GB 
disks (see http://www.kernel.org/ and see http://www.linuxdiskcert.org/). 
The latest stable kernel version is 2.4.20 (see http://www.kernel.org/). 
We needed the 2.4.x kernel to allow full support 
for ``Journaling'' file systems. Journaling file systems provide rapid 
recovery from crashes. A computer can finish 
its boot-up at a normal speed, rather than waiting to perform a file 
system check (FSCK) on the entire RAID array. This is then conducted 
in the background allowing the user to continue to use the RAID array. 
There are now 4 different Linux Journaling file systems: XFS, a port from 
SGI \cite{XFS}; JFS, a port from IBM \cite{JFS}; ext3 \cite{ext3}, 
a Journalized version of the standard ext2 file system; and ReiserFS 
from namesys \cite{ReiserFS}. Comparisons of these Journaling file 
systems have been done elsewhere \cite{Journaling}. 
When we tested our RAID-5 arrays only ext3 and the ReiserFS were 
easily available for the 2.4.x kernel; therefore, we tested 2 different 
Journaling file systems; ReiserFS and ext3. We opted on using ext3 for 
two reasons: 1) At the time there were stability problems with ReiserFS 
and NFS (this has since been resolved with kernel 
2.4.7) and 2) it was an extension of the standard ext2fs (it was 
originally developed for the 2.2 kernel) and, if synced properly could 
be mounted as ext2. Ext3 is the only one that will allow direct upgrading 
from ext2, this is why it is now the default for RedHat since 7.2.

NFS is a very flexible system that allows one to manage files on several 
computers inside a network as if they were on the local hard disk. So, 
there's no need to know what actual file system they are stored under nor 
where the files are physically located in order to access them. Therefore, 
we use NFS to connect these disks arrays to computers that cannot run 
Linux 2.4. We have successfully used NFS to mount disk arrays on the 
following types of computers: a DECstation 5000/150 running Ultrix 4.3A, 
a Sun UltraSparc 10 running Solaris 7, a Macintosh G3 running MacOS\,X, and 
various Linux boxes with both the 2.2 and 2.4 kernels. 

As an example, in Spring 2002 we built a pair of one Terabyte Linux RAID-5 arrays, 
as described in section \ref{hardware}, to store CMS Monte Carlo data at CERN. 
They were mounted using NFS, via gigabit ethernet. They remotely served the 
random background data to the CMS Monte Carlo Computers, as if it was local. 
While this is not as efficient as serving the data directly, it is clearly a viable 
technique \cite{CMS-Note}. We also are currently using two, NFS mounted, 
RAID-5 boxes, one at SLAC and one at the University of Mississippi, to run 
analysis software with the {\sc BaBar} KANGA and CMS CMSIM/ORCA code. 

We have performed a few simple speed tests. The first was \hbox{``hdparm 
-tT /dev/xxx''}. This test simply reads a 64 MB chunk of data and measures 
the speed. On a single drive we saw read/write speeds of about 30 MB/s. 
The whole array saw an increase to 95 MB/s. When we tried writing 
a text file using a simple FORTRAN program (we wrote ``All work and no play 
make Jack a dull boy'' $10^{8}$ times), the speed was about 95 MB/s While 
mounted via NFS over 100 Mb/s ethernet the speed was 2.12 MB/s, limited by 
both the ethernet speed and the NFS communication overhead. In the 
past \cite{farm}, we have been able to get much higher fractions of the rated 
ethernet bandwidth by using the lower level TCP/IP socket protocol \cite{TCPIP} 
in place of the higher level NFS protocol.  TCP/IP sockets are more cumbersome 
to program, but are much faster.

We also tested what actually happens when a disk fails by turning the power 
off to one disk in our RAID-5 array.  One could continue to read and write 
files, but in a ``degraded'' mode, that is without the parity safety net. 
When a blank disk was added to replace the failed disk, again one could 
continue to read and write files in a mode where the disk access speed is 
reduced while the system rebuilt the missing disk as a background job. 
This speed reduction in disk access was due to the fact that the parity 
regeneration is a major disk access in its own right. For more details, 
see reference \cite{HOWTO}.

The performance of Linux IDE software drivers is improving. The latest 
standards \cite{t13} include support for command overlap, READ/WRITE direct 
memory access QUEUED commands, scatter/gather data transfers without 
intervention of the CPU, and elevator seeks. Command overlap is a protocol 
that allows devices that require extended command time to perform a bus 
release so that commands may be executed by the other device on the bus. 
Command queuing allows the host to issue concurrent commands to the same 
device. Elevator seeks minimize disk head movement by optimizing the order 
of I/O commands. The Hitachi/IBM 180GXP disk \cite{IBM180} supports elevator 
seeks under the new ATA6 standard \cite{t13}.

We did encounter a few problems. We had to modify ``MAKEDEV'' to allow 
for more than eight IDE devices, that is to allow for disks beyond ``/dev/hdg''. 
For version 2.x one would have to actually modify the script; however, for 
version 3.x we just had to modify the file ``/etc/makedev.d/ide''. This should no 
longer be a problem with newer releases of Linux.

Another problem was the 2 GB file size limit. Older operating system and 
compiler libraries used a 32 bit ``long-integer'' for addressing files; 
therefore, they could not normally address files larger than 2 GB ($2^{31}$). 
There are patches to the Linux 2.4 kernel and glibc but there are still some 
problems with NFS and not all applications use these patches. 

We have found that the current underlying file systems (ext2, ext3, reiserfs) 
do not have a 2 GB file size limit.  The limit for ext2/ext3 is in the 
petabytes. The 2.4 kernel series supports large files (64-bit offsets). 
Current versions of GNU libc support large files. However, by default the 
32-bit offset interface is used. To use 64-bit offsets, C/C++ code must be 
recompiled with the following as the first line:
\begin{verbatim}
#define _FILE_OFFSET_BITS 64
\end{verbatim}
or the code must use the *64 functions (i.e. open becomes open64, etc.) 
if they exist. This functionality is not included in GNU FORTRAN (g77); 
however, it should be possible to write a simple wrapper C program to 
replace the OPEN statement (perhaps called open64). We 
have succeeded in writing files larger than 2 GB using a simple C 
program with ``\#define \_ FILE\_ OFFSET\_ BITS 64'' as the 
first line. This works over NFS version 3 but not version 2. 

While RAID-5 is recoverable for a hardware failure, there is no 
protection against accidental deletion of files. To address this 
problem we suggest a simple script to replace the ``rm'' command. 
Rather than deleting files it would move them to a ``/raid/Trash'' or 
better yet a ``/raid/.Trash'' directory on the RAID-5 disk array 
(similar to the ``Trash can'' in the Macintosh OS). The system 
administrator could later purge them as space is needed using an algorithm 
based on criteria such as file size, file age, and user quota.

\section{High Energy Physics Strategy}
\begin{figure}[ht!]
\includegraphics[width=65mm]{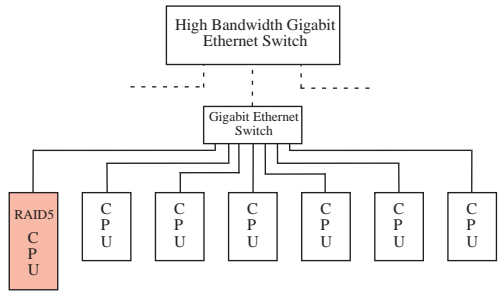}
\caption{
An example of a RAID-5 disk array mounted on several local CPUs via a 
8-port gigabit switch.}
\label{Cluster2}
\end{figure}

We encapsulate data and CPU processing power. A block of real or Monte 
Carlo simulated data for an analysis is broken up into groups 
of events and distributed once to a set of RAID disk boxes, which each 
may also serve a few additional processors via a local 8-port gigabit ethernet 
switch (see Figure \ref{Cluster2}).

Examples of commodity gigabit ethernet switches and PCI adapters are seen in 
Table \ref{Gigabit}.
\begin{table}[h!]
\begin{center}
\tabcolsep=1.5mm
\caption[]{Examples of Commodity Gigabit Ethernet Switches and Adapters.}
\label{Gigabit}
\begin{tabular}[t]{llcr}\hline
Company &Model&Type& Cost \\
\hline
Linksys \cite{Linksys} & EG008W &8-port switch& \$162 \\
D--Link \cite{Dlinkswitch} & DGS--1008T &8-port switch& \$312 \\
Netgear \cite{Netgear8} & GS508T &8-port switch& \$502 \\
Netgear \cite{Netgear24} & GS524T &24-port switch& \$1499 \\
D--Link \cite{Dlinkcard}  & DGE500T &PCI adapter & \$46 \\
Intel \cite{Intel}  & 82540EM &PCI adapter & \$41 \\
\hline
\end{tabular}
\end{center}
\end{table}
Dual processor boxes would also add more local CPU power. Events are stored 
on disks close to the CPUs that will process them to minimize I/O. Events are only 
moved once. Event parallel processing has a long history of success in high energy 
physics \cite{farm,E769,Kunz}. The data from each analysis are distributed among 
all the RAID arrays so all the computing power can be brought to bear 
on each analysis. 

For example, in the case of an important analysis (such 
as a Higgs analysis), one could put 50 GB of data onto each of 100 RAID arrays 
and then bring the full computing power of 700 CPUs into play. Instances of 
an analysis job are run on each local cluster in parallel. Several analyses 
jobs may be running in memory or queued to each local cluster to level loads. 
The data volume of the results (\eg ~histograms) is small and is gathered 
together over the network backbone. Results are examined and the analysis is 
rerun. The system is inherently fault tolerant. If three of a hundred clusters 
are down, one still gets 97\% of the data and analysis is not impeded. 

RAID-5 arrays should be treated as fairly secure, large, high-speed ``scratch 
disks''. RAID-5 just means that disk data will be lost less frequently. 
Data which is very hard to re-create still needs to reside on tape. The 
inefficiency of an offline tape vault can be an advantage. Its harder to erase 
your entire raw data set with a single keystroke, if thousands of tapes have 
to be physically mounted. Someone may ask why all the write protect switches 
are being reset before all is lost. Its the same reason the Air Force has real 
people with keys in ICBM silos.

The granularity offered by RAID-5 arrays allows a university or 
small experiment in a laboratory to set up a few terabyte computer farm, 
while allowing a large Analysis Site or Laboratory to set up a few 
hundred terabyte or a petabyte computer system. For a large site, they 
would not necessarily have to purchase the full system at once, but 
buy and install the system in smaller parts. This would have two 
advantages, primarily they would be able to spread the cost over a 
few years and secondly, given the rapid increase in both CPU power and 
disk size, one could get the best ``bang for the buck''.

\begin{figure*}[ht!]
\includegraphics[width=135mm]{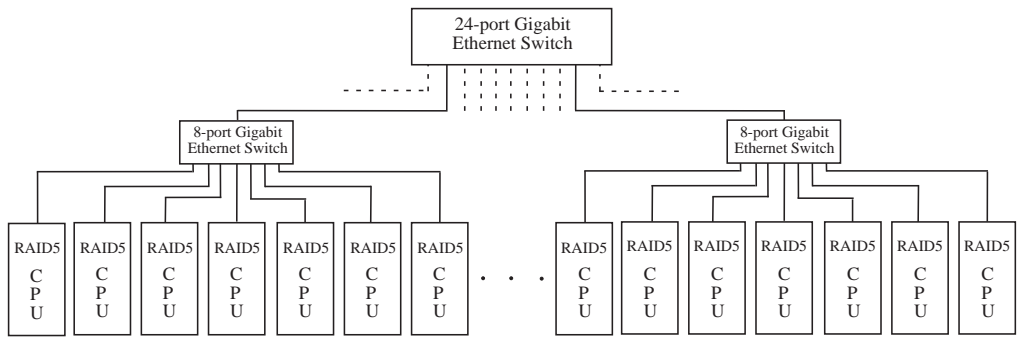}
\caption{
A schematic of a 1/4 petabyte  (or larger) system.}
\label{Cluster}
\end{figure*}

What would be required to build a 1/4 petabyte system (similar size  
as a tape silo)? Start with eight 250\,GB Maxtor disks in a box. The 
Promise Ultra133 card allows one to exceed the 137\,GB limit. 
Each box provides 
7 $\times $ 250\,GB = 1750\,GB of usable RAID-5 disk space in addition to 
a CPU for computations. 280 terabytes is reached with 161 boxes. Use 23 
commodity 8-port gigabit ethernet switches (\$170 each) to connect the 
161 boxes to a 24-port commodity gigabit ethernet switch.
See Figure \ref{Cluster}.
This could easily fit in a room that was formerly occupied by a few old 
Mainframes, say an area of about a hundred square meters. The power 
consumption would be 25 kilowatts, 45 kilowatts if they all start up at once. 
One would need to build up operational experience for smooth running.  
As newer disks arrive that hold yet more data, even a petabyte 
system would become feasible. If one still needed more processing power 
per RAID array you could substitute for each RAID-5 CPU shown in Figure 
\ref{Cluster},  6 CPUs plus 1 RAID-5 CPU connected by an 8-port gigabit 
ethernet switch as described in Figure \ref{Cluster2}. Multiple CPUs per 
motherboard provide another alternative to adjust the disk space to 
processing power ratio.

Grid Computing \cite{grid} will entail the movement of large amounts of 
data between various sites. RAID-5 arrays will be needed as disk caches 
both during the transfer and when it reaches its final destination. Another 
example that can apply to Grid Computing is the Fermilab Mass Storage 
System, Enstore \cite{enstore}, where RAID arrays are used as a disk 
cache for a Tape Silo. Enstore uses RAID arrays to stage tapes to disk 
allowing faster analysis of large data sets.

\section{Conclusion}
We have tested redundant arrays of IDE disk drives for use in 
offline high energy physics data analysis and Monte Carlo simulations. 
Parts costs of total systems using commodity IDE disks are now at 
the \$2000 per terabyte level, a lower cost per terabyte than Storage 
Technology tape silos. The disks, however, offer much better granularity; 
even small institutions can afford them. The faster access of disk versus 
tape is a major added bonus. We have tested software RAID-5 systems 
running under Linux 2.4 using Promise Ultra 133 disk controllers. RAID-5 
provides parity bits to protect data in case of a single catastrophic disk failure. 
Tape backup is not required for data that can be recreated with modest effort. 
Journaling file systems permit rapid recovery from crashes. Our data analysis 
strategy is to encapsulate data and CPU processing power. Data is stored on 
many PCs. Analysis for a particular part of a data set takes place locally on the 
PC where the data resides. The network is only used to put results together. 
Commodity 8-port gigabit ethernet switches combined with a single high end, 
fast backplane switch \cite{cisco} would allow one to connect over a thousand PCs, 
each with a terabyte of disk space. Some tasks may need more than one CPU 
to go through the data even on one RAID array.  For such tasks dual CPUs 
and/or several boxes on one local 8-port ethernet switch should be adequate 
and avoids overwhelming the backbone switching fabric connecting an entire 
installation.  Again the backbone is only used to put results together. 

Current high energy physics experiments, such as  {\sc BaBar} at SLAC, feature 
relatively low data acquisition rates, only 3 MB/s, less than a third of 
the rates taken at Fermilab fixed target experiments a decade ago 
\cite{farm}. The Large Hadron Collider experiments CMS and Atlas, 
with data acquisition rates starting at 100 MB/s, will be more challenging 
and require physical architectures that minimize helter skelter data movement 
if they are to fulfill their promise. In many cases, architectures designed 
to solve particular processing problems are far more cost effective than 
general solutions \cite{farm,E769}. 
As Steve Wolbers in his talk at CHEP03 \cite{Wolbers} reminded us, all 
data processing groups can not depend on Moore's Law to save them. 
Data acquisition groups want to write out additional interesting
events. Programmers like to adopt new languages that are further abstracted
from the CPUs running them.  Small objects and pointers seem to find their way
into code. Machines hate to interrupt pipelines and love direct addressing.
Universities want networks to transfers billions of events quickly. Even Gordon
Moore may not be able to do all of this simultaneously. Efficiency may still
be useful. Designing time critical code \cite{metcalf}, regardless of the
language chosen, to fit into larger blocks without pointers can increase speed
by a factor of 10 to 100. Code to methodically bit-pack events into the
minimum possible size may be worth writing \cite{bracker}. If events are
smaller, more can be stored on a given disk and more can be transferred over a
given network in a day. All of this requires planning at an early stage. No
software package will generate it automatically.

Techniques explored in this paper, physically encapsulating data and CPUs 
together, may be useful. Terabyte disk arrays at small institutions are now feasible. 
Computing has progressed since the days when science was done by punching 
a few kilobytes into paper tape \cite{lasker}.

\begin{acknowledgments}
Many thanks to S. Bracker, J. Izen, L. Lueking, R. Mount, M. Purohit, 
W. Toki, and T. Wildish for their help and suggestions.  
This work was supported in part by the U.S. Department of Energy under 
Grant Nos. DE-FG05-91ER40622 and DE-AC02-76CH03000.
\end{acknowledgments}


\vfill
\eject

\end{document}